\documentclass[12pt]{iopart}
\usepackage{graphicx}

\begin{document}

\title[Entanglement in open quantum systems]{Dynamics of quantum entanglement\\
in Gaussian open systems}

\author{Aurelian Isar}

\address{National Institute of Physics and Nuclear Engineering,
P.O.Box MG-6, Bucharest-Magurele, Romania}
\ead{isar@theory.nipne.ro}
\begin{abstract}
In the framework of the theory of open systems based on completely
positive quantum dynamical semigroups, we give a description of the dynamics of entanglement for a system consisting of two uncoupled harmonic oscillators interacting with a thermal environment. Using Peres-Simon necessary and sufficient criterion for separability of two-mode Gaussian states, we describe the evolution of entanglement in terms
of the covariance matrix for a Gaussian input state. For some values of the temperature of environment, the state keeps for all times its initial type:
separable or entangled. In other cases, entanglement generation, entanglement sudden death or a repeated collapse and revival of entanglement take place. We determine the asymptotic Gaussian maximally entangled mixed states (GMEMS) and their corresponding asymptotic maximal logarithmic negativity.

\end{abstract}


\section{Introduction}

Quantum entanglement of Gaussian states constitutes a fundamental
resource in continuous variable quantum information processing and communication \cite{bra1}. For the class of Gaussian states there exist necessary and sufficient criteria of entanglement \cite{sim,dua} and quantitative entanglement measures \cite{vid,gie}. The quantum information processing tasks are difficult to implement, due to the fact that any realistic quantum system is not isolated and it always interacts with its environment. Quantum coherence and entanglement of quantum systems are inevitably influenced during their interaction with the external environment. As a result of the irreversible and uncontrollable phenomenon of quantum decoherence, the purity and entanglement of quantum states are in most cases degraded. Therefore in order to describe realistically quantum information processes it is necessary to take decoherence and dissipation into consideration. Decoherence and dynamics of quantum entanglement in continuous variable open systems have been intensively studied in the last years \cite{oli,pra,ser1,ben1,avd,man,aphysa,aeur}.

When two systems are immersed in an environment, then, in addition to and at the same time with the quantum decoherence phenomenon, the environment can also generate a quantum entanglement of the two systems and therefore an additional mechanism to correlate them \cite{ben1,vvd1,ben2}. In this paper we study, in the framework of the theory of open systems based on completely positive quantum dynamical semigroups, the dynamics of the continuous variable entanglement of two identical harmonic oscillators coupled to a common thermal environment.  We are interested in discussing the correlation effect of the environment, therefore we assume that the two oscillators are uncoupled, i.e. they do not interact directly. The initial state of the subsystem is taken of Gaussian form and the evolution under the quantum dynamical semigroup assures the preservation in time of the Gaussian form of the state. In section 2 we write the Markovian master equation in the Heisenberg representation for two uncoupled harmonic oscillators interacting with a general environment and the evolution equation for the covariance matrix of the considered subsystem. By using the Peres-Simon criterion for separability of two-mode Gaussian states \cite{sim,per}, we investigate in section 3 the dynamics of entanglement for this system. For certain values of the environment temperature, the state keeps for all times its initial type: separable or entangled. For other values of the temperature, entanglement generation, entanglement sudden death or a repeated collapse and revival of entanglement take place. In recent papers, in a different model, Paz and Roncaglia \cite{paz1,paz2} studied numerically the exact entanglement behaviour of two identical oscillators in an infinite bath of other oscillators by using the exact master equation for quantum Brownian motion and showed that the entanglement can undergo three qualitatively different dynamical phases: sudden death, sudden death and revival, and no sudden death of entanglement. Finally, we determine the asymptotic Gaussian maximally entangled mixed states (GMEMS) and the corresponding asymptotic maximal logarithmic negativity, which characterizes the degree of entanglement of these states. A summary is given in section 4.

\section{Time evolution of covariance matrix for two harmonic oscillators}

We study the dynamics of the subsystem composed of two identical non-interacting oscillators in weak interaction with a thermal environment. In the axiomatic formalism based on completely positive quantum dynamical semigroups, the irreversible time evolution of an open system is described by the following general quantum Markovian master equation for an operator $A$ in the Heisenberg representation ($\dagger$ denotes Hermitian conjugation) \cite{lin,rev}:
\begin{eqnarray}{dA\over dt}={\rmi\over \hbar}[H,A]+{1\over
2\hbar}\sum_j(2V_j^{\dagger}AV_j - V_j^{\dagger}
V_jA-AV_j^{\dagger}V_j).\label{masteq}\end{eqnarray}
Here, $H$ denotes the Hamiltonian of the open system and the operators $V_j, V_j^\dagger,$ defined on the Hilbert space of $H,$ represent the interaction of the open system with the environment.

We are interested in the set of Gaussian states, therefore we introduce such quantum dynamical semigroups that preserve this set during time evolution of the system. Consequently $H$ is taken to be a polynomial of second degree in the coordinates $x,y$ and momenta $p_x,p_y$ of the oscillators and $V_j,V_j^{\dagger}$ are taken polynomials of first degree in these canonical observables. Then in the linear space spanned by coordinates and momenta there exist only four linearly independent operators $V_{j=1,2,3,4}$ \cite{san}: \begin{eqnarray}
V_j=a_{xj}p_x+a_{yj}p_y+b_{xj}x+b_{yj}y,\end{eqnarray} where
$a_{xj},a_{yj},b_{xj},b_{yj}$ are complex coefficients. The Hamiltonian $H$ of the two uncoupled identical harmonic oscillators of mass $m$ and frequency $\omega$ is given by \begin{eqnarray}
H=\frac{1}{2m}(p_x^2+p_y^2)+{m\omega^2\over
2}(x^2+y^2).\end{eqnarray}

The fact that the evolution is given by a dynamical semigroup
implies the positivity of the following matrix formed by the
scalar products of the four vectors $ {\bi a}_x, {\bi a}_y,
{\bi b}_x, {\bi b}_y$ whose entries are the components $a_{xj},a_{yj},b_{xj},b_{yj},$
respectively:
\begin{eqnarray}\frac{1}{2} \hbar \left(\matrix
{({\bi a}_x {\bi a}_x)&({\bi a}_x {\bi b}_x) &({\bi a}_x
{\bi a}_y)&({\bi a}_x {\bi b}_y) \cr ({\bi b}_x {\bi a}_x)&({\bi
b}_x {\bi b}_x) &({\bi b}_x {\bi a}_y)&({\bi b}_x {\bi b}_y)
\cr ({\bi a}_y {\bi a}_x)&({\bi a}_y {\bi b}_x) &({\bi a}_y
{\bi a}_y)&({\bi a}_y {\bi b}_y) \cr ({\bi b}_y {\bi
a}_x)&({\bi b}_y {\bi b}_x) &({\bi b}_y {\bi a}_y)&({\bi b}_y
{\bi b}_y}\right).
\end{eqnarray}
We take this matrix of the following form, where all diffusion coefficients $D_{xx}, D_{xp_x},$... and dissipation constant $\lambda$ are real quantities (we put from now on $\hbar=1$):
\begin{eqnarray} \left(\begin{matrix}{D_{xx}&- D_{xp_x} -\rmi \frac{\lambda}{2}&D_{xy}& -
D_{xp_y} \cr - D_{xp_x} +\rmi \frac{\lambda}{2}&D_{p_x p_x}&-
D_{yp_x}&D_{p_x p_y} \cr D_{xy}&- D_{y p_x}&D_{yy}&- D_{y p_y}
- \rmi \frac{\lambda}{2} \cr - D_{xp_y} &D_{p_x p_y}&- D_{yp_y} + \rmi
\frac{\lambda}{2}&D_{p_y p_y}}\end{matrix}\right).\label{coef} \end{eqnarray}
It follows that the principal minors of this matrix are positive or zero. From
the Cauchy-Schwarz inequality the following relations hold for the coefficients defined in  (\ref{coef}): \begin{eqnarray}
D_{xx}D_{p_xp_x}-D^2_{xp_x}\ge\frac{\lambda^2}{4},~~~D_{yy}D_{p_yp_y}-D^2_{yp_y}\ge\frac{\lambda^2}{4},\nonumber\\
D_{xx}D_{yy}-D^2_{xy}\ge0,~~~D_{p_xp_x}D_{p_yp_y}-D^2_{p_xp_y}\ge 0,\nonumber\\
D_{xx}D_{p_yp_y}-D^2_{xp_y}\ge 0,~~~D_{yy}D_{p_xp_x}-D^2_{yp_x}\ge 0.
\label{coefineq}\end{eqnarray}

A two-mode Gaussian state is completely characterized by its first and second moments of canonical variables.
We therefore introduce the following $4\times 4$ bimodal covariance matrix, which entirely specifies a two-mode Gaussian state (all first moments
have been set to zero by means of local unitary operations which do not affect the entanglement):
\begin{eqnarray}\sigma(t)=\left(\begin{matrix}{\sigma_{xx}(t)&\sigma_{xp_x}(t) &\sigma_{xy}(t)&
\sigma_{xp_y}(t)\cr \sigma_{xp_x}(t)&\sigma_{p_xp_x}(t)&\sigma_{yp_x}(t)
&\sigma_{p_xp_y}(t)\cr \sigma_{xy}(t)&\sigma_{yp_x}(t)&\sigma_{yy}(t)
&\sigma_{yp_y}(t)\cr \sigma_{xp_y}(t)&\sigma_{p_xp_y}(t)&\sigma_{yp_y}(t)
&\sigma_{p_yp_y}(t)}\end{matrix}\right).\label{covar} \end{eqnarray}
The problem of solving the master equation for the operators in Heisenberg representation can be transformed into a problem of solving first-order in time, coupled linear differential equations for the covariance matrix elements. Namely, from  (\ref{masteq}) we obtain the following system of equations for the quantum correlations of the canonical observables, written in matrix form \cite{san} ($\rm T$ denotes a transposed matrix):
\begin{eqnarray}{d \sigma(t)\over
dt} = Y \sigma(t) + \sigma(t) Y^{\rm T}+2 D,\label{vareq}\end{eqnarray} where
\begin{eqnarray} Y=\left(\begin{matrix} {-\lambda&1/m&0 &0\cr -m\omega^2&-\lambda&0&
0\cr 0&0&-\lambda&1/m \cr 0&0&-m\omega^2&-\lambda}\end{matrix}\right),\end{eqnarray}
\begin{eqnarray}D=\left(\begin{matrix}
{D_{xx}& D_{xp_x} &D_{xy}& D_{xp_y} \cr D_{xp_x}&D_{p_x p_x}&
D_{yp_x}&D_{p_x p_y} \cr D_{xy}& D_{y p_x}&D_{yy}& D_{y p_y}
\cr D_{xp_y} &D_{p_x p_y}& D_{yp_y} &D_{p_y p_y}} \end{matrix}\right).\end{eqnarray}
The time-dependent
solution of  Eq. (\ref{vareq}) is given by \cite{san}
\begin{eqnarray}\sigma(t)= M(t)[\sigma(0)-\sigma(\infty)] M^{\rm
T}(t)+\sigma(\infty),\label{covart}\end{eqnarray} where the matrix $M(t)=\exp(Yt)$ has to fulfill
the condition $\lim_{t\to\infty} M(t) = 0.$
In order that this limit exists, $Y$ must only have eigenvalues
with negative real parts. The values at infinity are obtained
from the equation \begin{eqnarray}
Y\sigma(\infty)+\sigma(\infty) Y^{\rm T}=-2 D.\label{covarinf}\end{eqnarray}

\section{Dynamics of two-mode continuous variable entanglement}

The characterization of the separability of continuous variable states using second-order moments of quadrature operators was given in Refs. \cite{sim,dua}. A two-mode Gaussian state is separable if and only if the partial transpose of its density matrix is non-negative [necessary and sufficient positive partial transpose (PPT) criterion]. A two-mode Gaussian state is entirely specified by its covariance matrix (\ref{covar}), which is a real, symmetric and positive matrix with the block structure
\begin{eqnarray}
\sigma(t)=\left(\begin{array}{cc}A&C\\
C^{\rm T}&B \end{array}\right),\label{cm}
\end{eqnarray}
where $A$, $B$ and $C$ are $2\times 2$ Hermitian matrices. $A$ and $B$ denote the symmetric covariance matrices for the individual one-mode states, while the matrix $C$ contains the cross-correlations between modes. Simon \cite{sim} derived a PPT criterion for bipartite Gaussian
continuous variable states: the necessary and sufficient criterion for separability is
$S(t)\ge 0,$ where \begin{eqnarray} S(t)\equiv\det A \det B+\left(\frac{1}{4} -|\det
C|\right)^2\nonumber\\- {\rm Tr}[AJCJBJC^{\rm T}J]- \frac{1}{4}(\det A+\det B)
\label{sim1}\end{eqnarray} and $J$ is the $2\times 2$ symplectic matrix
\begin{eqnarray}
J=\left(\begin{array}{cc}0&1\\
-1&0\end{array}\right).
\end{eqnarray}
Since the two oscillators are identical, it is natural to consider environments for which $D_{xx}=D_{yy},~ D_{xp_x}=D_{yp_y},~D_{p_xp_x}=D_{p_yp_y},~ D_{xp_y}=D_{yp_x}.$ Then both unimodal covariance matrices are equal, $A=B,$ and the entanglement matrix $C$ is symmetric.

\subsection{Time evolution of entanglement}

In order to describe the dynamics of entanglement, we use the PPT criterion \cite{sim,per} according to which a state is entangled if and only if the operation of partial transposition does not preserve its positivity. Concretely, we have to analyze the time evolution of the Simon function $S(t)$ (\ref{sim1}). For a thermal environment characterized by the temperature $T,$ we consider such diffusion coefficients, for which \begin{eqnarray}m\omega D_{xx}=\frac{D_{p_xp_x}}{m\omega}=
\frac{\lambda}{2}\coth\frac{\omega}{2kT},~~~D_{xp_x}=0,\label{envcoe1}\end{eqnarray}
\begin{eqnarray}m^2\omega^2D_{xy}=D_{p_xp_y}.\label{envcoe2}\end{eqnarray} This corresponds to the case when the asymptotic state is a Gibbs state \cite{rev}. We consider two cases, according to
the type of the initial Gaussian state: 1) separable and
2) entangled.

1) To illustrate a possible generation of the entanglement, we represent in Figure 1 the dependence
of function $S(t)$ on time $t$ and temperature $T$ for a separable initial Gaussian mixed state. We notice that,
according to Peres-Simon criterion, for relatively small
values of the temperature $T,$ the initial separable state
$(S(t) = 0)$ becomes entangled shortly after the initial moment of time $t = 0$. For relatively large
values of $T,$ $S(t)$ is strict positive and the state remains
separable for all times.

Depending on the environment temperature, there are three situations in the case of a generated entanglement \cite{arus1}:
a) entanglement may persist forever, including
the asymptotic final state; b) there exist repeated collapse and revival of entanglement; c) the entanglement is
created only for a short time, then it disappears and the
state becomes again separable. The entanglement of
the two modes can be generated from an initial separable state during the interaction with the environment
only for certain values of mixed diffusion coefficient $D_{xp_y}$
and dissipation constant $\lambda.$

2) An example of the evolution of an entangled initial state is illustrated in Figure 2, where we represent the
dependence of function $S(t)$ on time $t$ and temperature $T$ for
an entangled initial Gaussian mixed state.
For relatively small values of $T,$ the initial entangled state
may remain entangled for all times \cite{arus1}. For relatively large values of temperature $T,$ at some finite moment of time,
$S(t)$ takes non-negative values and therefore the state
becomes separable. This is the so-called phenomenon of
entanglement sudden death. This phenomenon is in contrast to the loss of quantum coherence, which is usually
gradual [12]. Depending on the values of the temperature, it is also possible to have a repeated collapse and
revival of the entanglement.

\begin{figure}
{
\includegraphics{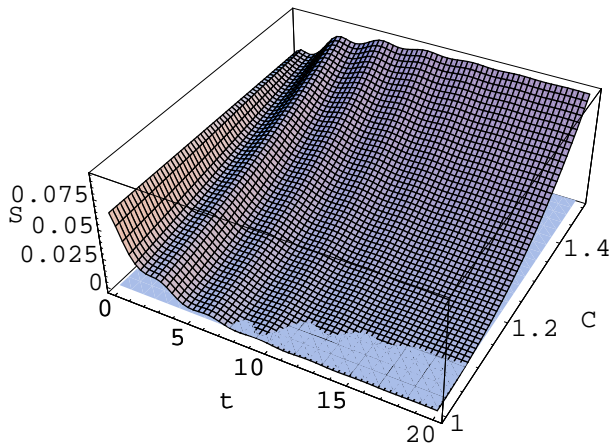}}
\caption{Dependence of Simon separability function $S$ on time $t$
and environment temperature $C\equiv\coth\frac{\hbar\omega}{2kT}$ for $\lambda=0.1,$ $D_{xp_y}=0.049$ and for a separable initial Gaussian mixed state with initial correlations $\sigma_{xx}(0)=1,~\sigma_{p_xp_x}(0)=1/2,~\sigma_{xp_x}(0)=\sigma_{xy}(0)=\sigma_{p_xp_y}(0)=
~\sigma_{xp_y}(0)=0.$ We take $m=\omega=\hbar=1.$
}
\label{fig:1}
\end{figure}

\begin{figure}
{
\includegraphics{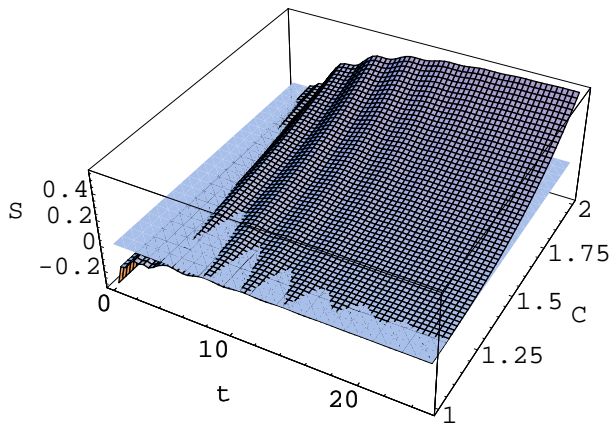}}
\caption{Same as in Figure 1, for an entangled initial Gaussian mixed state with initial correlations $\sigma_{xx}(0)=1,~\sigma_{p_xp_x}(0)=1/2,~\sigma_{xp_x}(0)=0,~\sigma_{xy}(0)=1/2, ~\sigma_{p_xp_y}(0)=-1/2,~\sigma_{xp_y}(0)=0.$
}
\label{fig:2}
\end{figure}

\subsection{Logarithmic negativity}

For Gaussian states, the measures of entanglement of bipartite systems are based on some invariants constructed from the elements of the covariance matrix \cite{oli,avd}. Heisenberg uncertainty principle can be expressed as a constraint on the global symplectic
invariants $\Delta \equiv \det A + \det B + 2\det C$
and $\det \sigma$ \cite{sim}:
\begin{eqnarray}
\Delta\le\frac{1}{4}
+ 4\det \sigma, \label{heis}\end{eqnarray}
and the symplectic eigenvalues
$\nu_{\mp},$ which form the symplectic spectrum of covariance matrix $\sigma$
are determined by
\begin{eqnarray}
2\nu_{\mp}^2=\Delta\mp\sqrt{\Delta^2-4\det\sigma}.
\label{eigs}\end{eqnarray}
In terms of $\nu_{\mp}$ relation (\ref{heis}) takes the form $\nu_{-}\ge1/2$.

In order to quantify the degrees of entanglement of the infinite-dimensional bipartite system states of the two oscillators it is suitable to use the logarithmic negativity.  For a Gaussian density operator, the logarithmic negativity is completely defined by the symplectic spectrum of the partial transpose of the covariance matrix. It is given by
$
E_N={\rm max}\{0,-\log_2 2\tilde\nu_-\},
$
where $\tilde\nu_-$ is the smallest of the two symplectic eigenvalues of the partial transpose $\tilde{{\sigma}}$ of the 2-mode covariance matrix $\sigma:$
\begin{eqnarray}2\tilde{\nu}_{\mp}^2 = \tilde{\Delta}\mp\sqrt{\tilde{\Delta}^2
-4\det\sigma}.
\end{eqnarray}
Here $ \tilde\Delta$ is given by
$ \tilde\Delta=\Delta - 4\det C = \det A+\det B-2\det C.$ A state is separable
if and only if \begin{eqnarray}\tilde\nu_{-}\ge1/2\label{seigen}\end{eqnarray} and logarithmic negativity quantifies the violation of inequality (\ref{seigen}).

In our model, the logarithmic negativity is calculated as \begin{eqnarray}E_N(t)=-\frac{1}{2}\log_2[4f(\sigma(t))], \end{eqnarray} where \begin{eqnarray}f(\sigma(t))=\frac{1}{2}(\det A +\det
B)-\det C\nonumber\\
-\left({\left[\frac{1}{2}(\det A+\det B)-\det
C\right]^2-\det\sigma(t)}\right)^{1/2}.\end{eqnarray}
It determines the strength of entanglement for $E_N(t)>0,$ and if $E_N(t)\le 0,$ then the state is
separable.

As expected, the logarithmic negativity has a behaviour similar to that one of the Simon function in what concerns the characteristics of the state of being separable or entangled \cite{arus,aijqi,ascri,aosid}.

\subsection{Asymptotic entanglement}

From (\ref{covarinf}), (\ref{envcoe1}) and (\ref{envcoe2}) we obtain the following elements of the asymptotic matrices $A(\infty)=B(\infty):$
\begin{eqnarray} m\omega\sigma_{xx}(\infty)=\frac{\sigma_{p_xp_x}(\infty)}{m\omega}=\frac{1}{2}\coth\frac{\omega}{2kT}, ~~~\sigma_{xp_x}(\infty)=0
\label{varinf} \end{eqnarray}
and of the entanglement matrix $C(\infty):$
\begin{eqnarray}\sigma_{xy} (\infty) =
\frac{m^2(\lambda^2+\omega^2)D_{xy}+m\lambda
D_{xp_y}}{m^2\lambda(\lambda^2+\omega^2)},\end{eqnarray}
\begin{eqnarray}\sigma_{xp_y}(\infty)=
\sigma_{yp_x}(\infty)=\frac{\lambda
D_{xp_y}}{\lambda^2+\omega^2},\end{eqnarray}
\begin{eqnarray}\sigma_{p_xp_y} (\infty) =
\frac{m^2\omega^2(\lambda^2+\omega^2)D_{xy}-m\omega^2\lambda D_{xp_y}}{\lambda(\lambda^2+\omega^2)}.\end{eqnarray}

The mixedness of a quantum state $\rho$ is characterized by its
purity $\mu\equiv {\rm Tr}\rho^2$. For a two-mode
Gaussian state
the purity is given by
$\mu=1/4\sqrt{\det\sigma}$. The marginal purities $\mu_i$  ($i=1,2$) of the reduced states in mode $i$ are given by
$\mu_1=1/2\sqrt{\det A}$ and $\mu_2=1/2\sqrt{\det B}.$ The global and marginal purities range from $0$ to $1$,
and they fulfill the constraint $\mu \ge \mu_1 \mu_2,$ as a direct consequence of Heisenberg uncertainty relations. In Ref. \cite{ade1} the following
upper and lower bounds on the invariant $\Delta$ have been obtained in terms of global and marginal purities:
\begin{eqnarray}
\frac{1}{2 \mu} + \frac{(\mu_1 - \mu_2)^2}{4\mu_1^2 \mu_2^2}\le\Delta
\le \nonumber\\
\le \min \left\{ \frac{(\mu_1 + \mu_2)^2}{4 \mu_1^2 \mu_2^2}
- \frac{1}{2 \mu}, \frac{1}{4} \left(1+\frac{1}{\mu^2}\right)  \right\}.\label{ineqd}
\end{eqnarray}
The invariant $\Delta$ has a direct physical
interpretation \cite{ade1}: at given global and marginal
purities, $\Delta$ determines the amount of entanglement
of the state and the smallest symplectic eigenvalue $\tilde\nu_-$ of the partially transposed state
is strictly monotone in $\Delta.$ Consequently, the entanglement of a
Gaussian state with fixed global purity $\mu$ and marginal
purities $\mu_1,\mu_2$ is strictly increasing with decreasing $\Delta$.
According to double inequality (\ref{ineqd}), giving lower and upper
bounds on $\Delta,$ there exist both maximally and minimally
entangled Gaussian states.

We analyze the existence of the entanglement in the asymptotic regime in the symmetric situation $A=B.$ First we consider the particular case $D_{xy}=0.$ In the limit of long times, we obtain the following quantities:
\begin{eqnarray}
\det A = \det B = \frac{C^2_T}{4},~\det C = - \frac {d^2}{\Lambda^2}, ~
\det\sigma = \left(\frac{C^2_T}{4} - \frac{d^2}{\Lambda^2}\right)^2,\end{eqnarray}
where we used the notations:
\begin{eqnarray}
C_T \equiv \coth\frac {\omega}{2kT},~ d \equiv D_{xp_y},~ \Lambda^2 \equiv \omega^2 + \lambda^2.\end{eqnarray}
Then we obtain
\begin{eqnarray}
\frac{1}{\mu}=C^2_T - 4\frac{d^2}{\Lambda^2},~\frac{1}{\mu_1} =\frac{1}{ \mu_2} = {C_T},~\Delta = 2\sqrt{\det\sigma}=\frac{1}{2\mu}.
\end{eqnarray}
These values saturate the lower bound in inequalities (\ref{ineqd}) and this situation entails a maximal entanglement. Consequently, the corresponding states are Gaussian maximally entangled mixed states (GMEMS). They are thermal squeezed states with the squeezing parameter $r$ given by $\tanh 2r = d/\Lambda C_T. $ In the pure case ($\det \sigma = 1/16$), these states are equivalent to two-mode squeezed vacua with the squeezing parameter determined only by the temperature of the thermal environment :
$\tanh 2r = \sqrt{C^2_T-1}/2C_T. $

According to Ref. \cite{ade1},  these states are separable in the range
\begin{equation}
\mu \le \frac{\mu_1 \mu_2}{\mu_1 + \mu_2 - \mu_1 \mu_2}.\label{sepreg}
\end{equation}
Then, for a given temperature $T,$ we obtain that the asymptotic final state is separable for the following range of positive values of the mixed diffusion coefficient $d$:
\begin{eqnarray}
\frac{2d}{\Lambda}\le C_T-1.\label{sep}\end{eqnarray} We remind that, according to inequalities (\ref{coefineq}), the coefficients have to fulfill also the constraint $\lambda C_T/2 \ge d.$
For a given temperature of the environment and for this range of mixed diffusion coefficients,
no entanglement can occur for these states. Outside this region, i.e. for a temperature and diffusion coefficient
satisfying \begin{eqnarray}
C_T-1\le\frac{2d}{\Lambda}\le C_T+1,\label{entan}\end{eqnarray}
they are Gaussian maximally entangled mixed states (GMEMS).

The asymptotic logarithmic negativity has the form
\begin{eqnarray} E_N(\infty)=-\log_2\left(C_T-\frac{2d}{\Lambda}\right).\end{eqnarray}
Outside the separable region, this is the maximum possible value of the
logarithmic negativity, attained by GMEMS.
It depends only on the mixed diffusion coefficient, dissipation constant and temperature, and does not depend on the initial Gaussian state.

When $d=0,$ but $D_{xy}\neq 0,$ then $\det C> 0$ and the asymptotic final state is always separable. If both these diffusion coefficients are non-zero, $d\neq 0,$ $D_{xy}\neq 0,$ then the lower bound in inequalities (\ref{ineqd}) is not anymore saturated and, consequently, the corresponding states are Gaussian non-maximally entangled mixed states.

We notice that the asymptotic persistence of entanglement is the result of the competition between thermal decoherence, determined by the uni-modal diffusion coefficients (\ref{envcoe1}) and the statistical coupling of the two modes, due to the presence of position-position and momentum-momentum diffusion coefficients, which are chosen of equal relevance in our model ($m^2\omega^2D_{xy}=D_{p_xp_y}$) and position-momentum diffusion coefficients $D_{xp_y}=D_{yp_x}\equiv d.$

\section{Summary}

In the framework of the theory of open quantum systems based on completely positive quantum dynamical semigroups, we investigated the Markovian dynamics of the quantum entanglement for a subsystem composed of two noninteracting modes embedded in a thermal environment. By using the Peres-Simon necessary and sufficient criterion for separability of two-mode Gaussian states, we have described the evolution of entanglement in terms of the covariance matrix for Gaussian input states. The dynamics of the quantum entanglement is sensitive to the initial states and the parameters characterizing the environment (diffusion and dissipation coefficients and temperature). For some values of these parameters, the state keeps for all times its initial type: separable or entangled. In other cases, entanglement generation or entanglement suppression (entanglement sudden death) take place or even one can notice repeated collapse and revival of entanglement. We have also shown that, independent of the type of the initial state - separable or entangled, for certain values of temperature, the initial state evolves asymptotically to an equilibrium state which is entangled, while for other values of temperature the asymptotic state is separable. For a given temperature, we calculated the range of mixed diffusion coefficients which determine the existence of asymptotic Gaussian maximally entangled mixed states (GMEMS). We obtained also the expression of the maximal logarithmic negativity, which characterizes the degree of entanglement of these states.

\ack
The author acknowledges the financial support received within
the Project CNCSIS-IDEI 497/2009 and Project PN 09 37 01 02/2009. The author thanks Victor Dodonov and
Salomon Mizrahi for invitation and kind hospitality.

\end{document}